\documentclass[twocolumn,aps,prd]{revtex4}
\usepackage{graphicx}
\usepackage{amsmath,amssymb}
\usepackage{color}
\usepackage[latin1]{inputenc}
\usepackage[english]{babel}

\newcommand{\GeV}{\text{ GeV}}
\newcommand{\TeV}{\text{ TeV}}
\newcommand{\seconds}{\text{ s}}

\newcommand{\years}{\text{ y}}
 
\newcommand{\mgrav}{m_{3/2}}

\newcommand{\mplanck}{\ensuremath{m_{\text{pl}}}}

\def\lsim{\mathrel{\rlap{\lower4pt\hbox{\hskip1pt$\sim$}}
 \raise1pt\hbox{$<$}}}                
\def\gsim{\mathrel{\rlap{\lower4pt\hbox{\hskip1pt$\sim$}}
 \raise1pt\hbox{$>$}}}                

\begin{document} 

\title{Testing Superstring Theories with Gravitational Waves\\
}

\author{{Ruth Durrer$^a$ and Jasper Hasenkamp$^b$}\\[0.5cm]
{\it
a Department of Theoretical Physics and Center for Astroparticle Physics, University of Geneva, Geneva, Switzerland \\ and CEA, SPhT, URA 2306, F-91191 Gif-sur-Yvette, France{\tiny, (email:Ruth.Durrer@unige.ch)}}\\
{\it
b II. Institute for Theoretical Physics, University of Hamburg, Hamburg, Germany{\tiny, (email:Jasper.Hasenkamp@desy.de)}}\\
[0.15cm]
}

\begin{abstract} 
We provide a simple transfer function that determines the effect of an early 
matter-dominated era on the gravitational wave background
and show that a large class of compactifications of superstring theory
might be tested by observations of the gravitational wave background from inflation.
For large enough reheating temperatures $\gtrsim 10^9 \GeV$
the test applies to all models containing at least one scalar with 
mass $\lesssim 10^{12}\GeV$ that acquires a large initial oscillation amplitude after inflation 
and has only gravitational interaction strength, i.e., a field with the typical properties 
of a modulus.
\end{abstract}


\maketitle


\section{Introduction}
Our description of the forces of Nature consisting of the 
standard model of particle physics and general relativity as the theory of gravity
is astonishingly successful.
However, the theory is incomplete.
On large scales gravitational dynamics requires the introduction of 'invisible' 
components like dark matter and dark energy.
The standard model 
of particle physics is lacking neutrino masses and contains many unexplained parameters.
The most serious problem, however, is 
that we have no quantum theory of gravity. 
Classical general relativity has singularities. It breaks 
down when the curvature becomes too large.
The most developed attempt for a quantum 
theory of gravity is string theory which may turn out to be 
the fundamental theory of Nature.

String theory has the weakness that it is very difficult to test experimentally. This is not only 
due to the fact that 'stringy' effects become relevant only at very high energy, but is also  
a consequence of the
landscape~\cite{Becker:1996gj}. It has turned out that string models
are extremely versatile and able to predict more or less everything. However, 
if we want to take some version of string theory seriously as a physical theory of Nature, it needs to be falsifiable. 
In this paper we propose a test for the existence
of at least one scalar field in the model that 
affects the cosmic evolution after inflation.
The well-known moduli problem~\cite{Coughlan:1983ci}
predicts a phase of matter dominated expansion some time after inflation. This matter
has either to be diluted by a subsequent phase of 'thermal 
inflation'~\cite{Coughlan:1984yk} or, more naturally,  the moduli have to decay.
In this paper we concentrate on this second possibility which does not require 
any additional ingredients. 
Qualitatively, however, our discussion also holds for thermal inflation~\cite{Easther:2008sx}.

Moduli fields describe the configuration of the curled up
extra dimensions of the string compactification.
They have gravitational coupling strength only
which makes them long-lived. They must be stabilized in order for the
measured masses and couplings of standard model particles to have well-defined 
constant values as observed. Therefore, they must have finite mass. 
In superstring models their mass $m_\phi$
is typically of the order of the gravitino mass $\mgrav$, 
so that  their decay width is $\Gamma \sim m_\phi^3/\mplanck^2 \sim \mgrav^3/\mplanck^2$,
where $\mplanck$ denotes the Planck mass.
If they are displaced from the origin after inflation,
they perform oscillations
and the Universe becomes matter dominated soon after.

If the moduli were cosmologically stable they would overclose the Universe and 
if they would decay after $t_\text{BBN} \sim 0.1 \seconds$ 
the success of big bang nucleosynthesis (BBN) were spoiled, because neutrinos 
would not have had enough time to thermalize~\cite{Kawasaki:1999na}.  Altogether, 
the moduli must decay before BBN.

We show that such a matter dominated phase before BBN leaves a detectable imprint on the
gravitational wave background from inflation: it significantly reduces the amplitude of the 
gravitational wave background at frequencies accessible to ground- or space-based 
detectors, compared to those probed by observations of the cosmic microwave background (CMB). 
Such a spectrum could be ruled out if a  gravitational wave background
from inflation were detected not only by future CMB experiments, but also by gravitational 
wave experiments at higher frequencies with an unsuppressed amplitude.
In this paper we calculate this signature in detail.

To circumvent the proposed test 
the modulus would have to decay before $t \sim 10^{-22}\seconds \simeq \Gamma^{-1}$
which corresponds to  $m_\phi = (\mplanck^2\Gamma)^{1/3} \sim 10^{12} \GeV$.

A particularly interesting situation arises for
superstring models with stabilized moduli that
have at least one long lived modulus
or modulus-like field whose mass is less than,
or of the order of, the gravitino mass~\cite{Kane:2010zz,Acharya:2010af}.
As mentioned above, the proposed test applies up
to scalar masses of $10^{12} \GeV$. 
Even though the tested mass range is well below
the Planck scale, this would be a crucial test,
because to our knowledge there is no other test for so massive, shortlived particles
which are way beyond being testable in colliders, also in high precision tests.
In this case the gravitino masses up to
the same order of magnitude, $10^{12} \GeV$, are probed.
As far as we know, no other possibility has been
proposed to probe so high supersymmetry breaking scales, albeit indirectly. 

Inflation does not only solve the horizon and flatness problems, but it also generates
a scale invariant spectrum of scalar and tensor (gravitational-waves) fluctuations,
see e.g.~\cite{Durrer:2008aa}.
Even though we know that the Universe was radiation
dominated at BBN, its evolution history before that is unknown.

In the following we examine the impact
of a moduli dominated phase  on the inflationary gravitational-wave background
and discuss the prospects for observations
in the inflationary gravitational-wave background.
\vspace{0.1cm}

\section{Evolution of gravitational waves from inflation}
Gravitational waves are the tensor perturbations $h_{ij}$ of 
the space-time metric,
\begin{equation}
ds^2 = a^2(\eta) (-d\eta^2 + (\delta_{ij} + 2 h_{ij}) dx^i dx^j )\,,
\end{equation} 
where $a(\eta)$ denotes the cosmic scale factor. The perturbation $h_{ij}$ is traceless, 
$h_i^i=0$, and divergence free, $\partial^i h_{ij} =0$.
The conformal time $\eta$ is defined by $d \eta = dt/a(t)$, where
$t$ denotes the physical time.
The energy density of gravitational waves is then given 
by~\cite{Weinberg:1972aa,Caprini:2001nb}
\begin{equation}
\rho_\text{gw}(\mathbf{x},t) = \frac{\langle \dot{h}_{ij}(\mathbf{x},t) \dot{h}^{ij}(\mathbf{x},t) 
\rangle}{8 \pi G a^2} \, ,
\label{rhogw}
\end{equation} 
where $G$ is Newton's constant. In this work an overdot indicates 
the derivative w.r.t conformal time $\eta$.
In Fourier space, the evolution of a gravitational-wave mode $h$ in a 
Friedmann universe (neglecting anisotropic stresses) is determined by~\cite{Durrer:2008aa}
\begin{equation}
\ddot{h} + 2 \frac{\dot{a}}{a} \dot{h} + k^2 h =0 \, .
\label{gwevolution}
\end{equation} 
Introducing $x=k\eta$, and assuming power law 
expansion, $a\propto \eta^q$, this equation has the simple general solution
\begin{equation}
h =  \frac{x}{a(\eta)} \left(c_1 j_{q-1}(x) + c_2 y_{q-1}(x) \right) \, ,
\label{gensol}
\end{equation} 
where $j_n$ and $y_n$ denote the spherical Bessel functions of order $n$ as defined, 
e.g., in~\cite{Abramowitz:1972}.  One might replace $x/a$ by $x^{1-q}$ and adjust 
the pre-factors correspondingly. With this replacement it becomes evident that on super-Hubble 
scales, $x<1$, the $j$-mode
is constant while the $y$-mode behaves as $x^{-2q+1}$. From this general solution together
with Eq.~\eqref{rhogw} one infers that $\rho_\text{gw} \propto a^{-4}$ as soon as 
the wavelength is sub-Hubble, $x>1$. (On super Hubble scales the 'energy density' of a mode is not a
meaningful concept.) 

It is reasonable to assume that both modes have similar amplitudes after inflation, 
where $x\ll 1$ for all modes of interest. If $q>1/2$, the  $y$-mode is decaying and soon 
after inflation we may approximate the solution by the $j$-mode.
Note that a constant value of $q$ corresponds to a constant background equation of state
with the ratio of pressure $P$ to energy density $\rho$ given by 
$$w=P/\rho\qquad \mbox{and} \qquad q=2/(3w+1)\,. $$
For a non-inflating $(3w+1>0)$ universe, $q\ge1/2$
corresponds to $w\le1$ and comprises all  cases of interest. During inflation $-1/3>w\gsim -1$ and $q \lsim -1$.

\begin{figure}[t]
\centering
\includegraphics[width=\columnwidth]{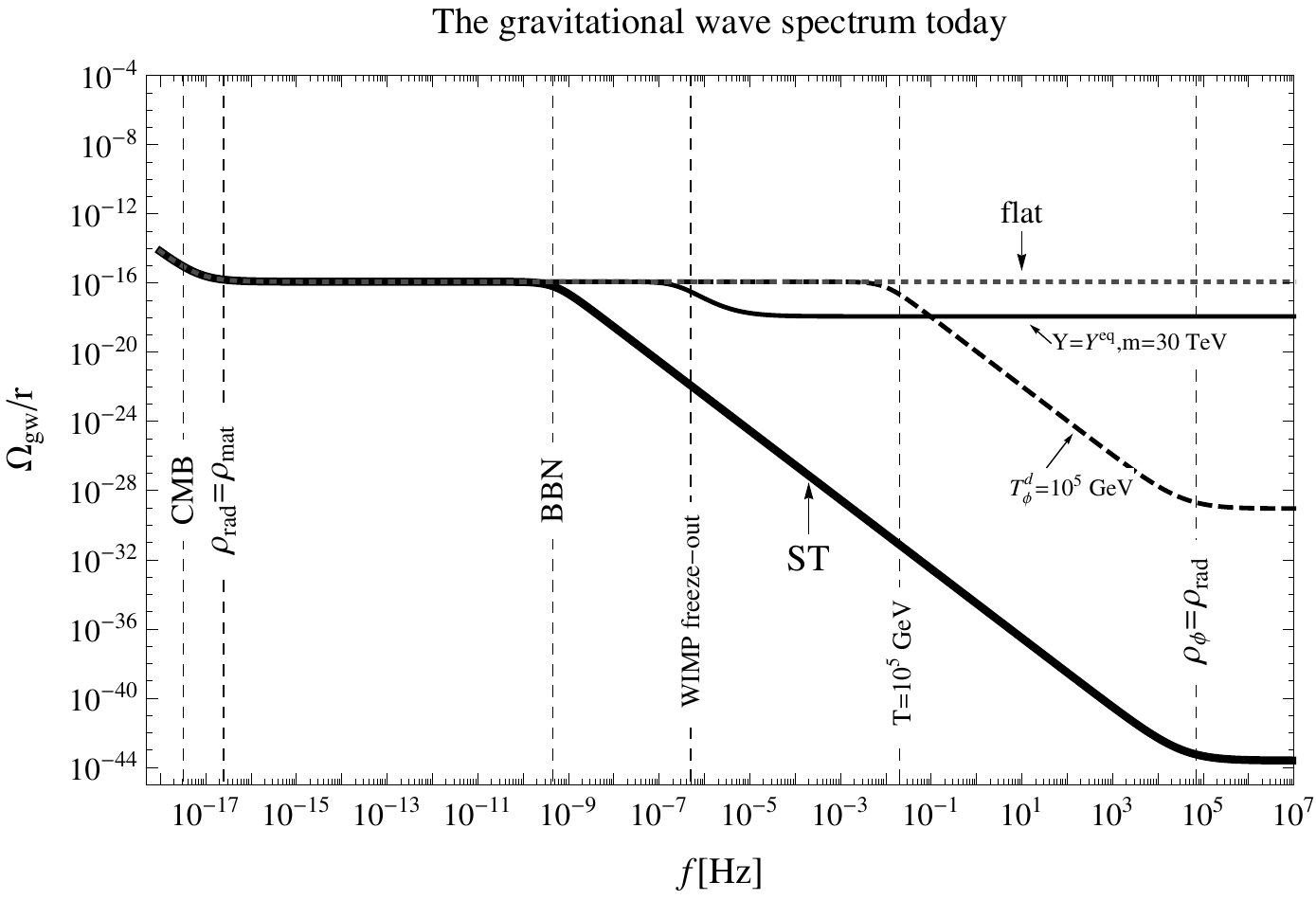} 
\caption{
Intensity of inflationary gravitational waves vs. their frequency $f= k/(2\pi)$ observed today.
The thick solid line shows the expectation from the usual cosmological moduli problem. 
The dashed line corresponds to the case of a much earlier modulus decay at 
$T_\phi^\text{d}=10^6 \GeV$. For comparison the dotted line shows a perfectly flat 
spectrum. The thin solid line demonstrates the impact on the spectrum of a particle 
with $30 \TeV$ mass that enters thermal equilibrium after inflation and decays before 
WIMP freeze-out. Various frequencies corresponding to important
and suggestive scales are highlighted by the vertical dashed lines:
CMB indicates the scale of best sensitivity of CMB experiments, and the other frequencies
relate to the horizon scale at the indicated event.
}
\label{fig:gwspectra}
\end{figure}

In standard cosmology, the Universe is radiation dominated after reheating until the time of
equality and matter dominated afterwards. Therefore $q=1$ until equality, where 
$\rho_\text{rad} = \rho_\text{mat}$, and $q=2$ after that. 
Since the energy density in gravitational waves scales like radiation, its fraction is constant on
scales which enter the horizon during the radiation dominated era and scales like 
$a(\eta_k) \propto \eta_k^2 \propto 1/k^2$ for scales which enter during the matter 
dominated era. A good approximation to the transfer function $T^2_{\rm eq}(k)$ which relates the
energy density per logarithmic $k$-interval to the amplitude of the gravitational-
wave spectrum after inflation in standard cosmology is 
given in~\cite{Turner:1993vb}. With this we obtain (for simplicity we neglect changes in 
the number of effective degrees of freedom and the minor effect of today's vacuum domination)
\begin{equation}
\Omega_\text{gw}(k) \equiv \frac{1}{\rho_\text{c}} \frac{d \rho_\text{gw}(k)}{d \text{log}(k)} =
\Omega_\text{rad} \frac{r \Delta_\mathcal{R}^2}{12 \pi^2} T_\text{eq}^2(k)\, ,
\label{gwspec}
\end{equation} 
where 
\begin{equation}
T_\text{eq}^2(k)= (1+ 1.57 \eta_\text{eq} k +3.42 (\eta_\text{eq} k)^2) (\eta_\text{cmb} k)^{-2}\,.
\label{Teq}
\end{equation} 
Here $ \Delta_\mathcal{R}$ is the amplitude of 
density fluctuations from 
inflation as measured in the CMB by the WMAP 
experiment~\cite{Komatsu:2008hk}, $ \Delta_\mathcal{R}^2\simeq 2\times 10^{-9} $,
and $\Omega_\text{rad} \simeq 5 \times 10^{-5} $.
The ratio $r$ is the tensor to scalar ratio which depends on the inflationary model,  
$\eta_\text{eq}$ and $\eta_\text{cmb}$ are the conformal time at matter-radiation equality 
and at CMB decoupling  respectively.
This standard spectrum is indicated by the dotted line in Fig.~\ref{fig:gwspectra}.
For Eq.~(\ref{Teq}) a Harrison-Zel'dovich spectrum is assumed. For different primordial 
spectra with spectral index $n_s\neq 1$ and $n_T\neq 0$ of the primordial scalar and tensor
fluctuations from inflation, the result (\ref{Teq}) has to be multiplied by $(k/k_c)^{n_T}$
if the amplitude $\Delta_\mathcal{R}$ and $r$ (which then are scale dependent) are 
determined at the pivot scale $k_c$.   
Changes in the background by nonstandard evolution of the Universe have previously been studied in~\cite{Sahni:1990tx}.

We now consider the situation that is likely to occur in superstring models, where 
soon after inflation
moduli come to dominate when $\rho_\phi=\rho_\text{rad}$. We denote the corresponding (conformal) time by
$\eta_\text{b}$. We assume that the moduli  decay briefly before nucleosynthesis
at time $\eta_\text{e}$. We then compute the final gravitational-wave spectrum by
matching the radiation solution ($q=1$) before $\eta_\text{b}$ to the matter solution ($q=2$) at 
$\eta_\text{b}$ and back to the radiation solution at $\eta_\text{e}$. After $\eta_\text{e}$  the 
Universe follows the standard evolution, so that the resulting spectrum simply has 
to be multiplied by the standard transfer function $T_\text{eq}^2(k)$.  The generic
shape of the resulting transfer function $T$ is clear from the general solution:
On super-Hubble scales the solution remains constant and $T=1$. Scales that enter 
the horizon during the matter dominated phase at $\eta_b<\eta_k=1/k< \eta_e$ are 
suppressed by a factor $a(\eta_k)/a(\eta_e) = (k\eta_e)^{-2}$ since 
$\rho_\text{gw} \propto a^{-4}$ while $\rho_\text{mat} \propto a^{-3}$. Scales which have already 
entered before matter domination are maximally suppressed by a factor
$a(\eta_\text{b})/a(\eta_\text{e}) = (\eta_\text{b}/\eta_\text{e})^{2}$. 

For sufficiently long matter domination, $\eta_\text{e}/\eta_\text{b} \ge 4$, we find
the following simple and accurate analytic approximation to 
the exact result for the transfer function
of an intermediate matter dominated phase:
\begin{equation}
T^2(k; \eta_e,\eta_b) \simeq \frac{1}{
\frac{\eta_\text{e}^2}{\eta_\text{b}^2} 
\left(\frac{2\pi c}{k\eta_b}-\frac{2\pi }{k\eta_e} +1\right)^{-2}
+1} \, ,
\label{tsquare}
\end{equation} 
where the best-fit analysis gives $c=0.5$.
The currently observable gravitational-wave spectrum is then simply
\begin{equation}
\Omega_\text{gw}(k)  =
\Omega_\text{rad} \frac{r \Delta_\mathcal{R}^2}{12 \pi^2} T_\text{eq}^2(k) T^2(k; \eta_e,\eta_b) \, ,
\label{gwspecfinal}
\end{equation}
with the fitting formula for $T(k; \eta_e,\eta_b)$ from Eq.~(\ref{tsquare}).
The time when matter domination begins, $\eta_\text{b}$, is
determined by the mass of the modulus field. 
Up to a coupling constant of order one, the mass also determines
the time of the modulus decay, $\eta_\text{e}$, which is the time when matter domination
ends.
Using the general formula~\cite{Durrer:2010xc} for conformal time,
\begin{equation}
\eta = 1.5\times 10^5 \seconds \left(\frac{100\text{ GeV}}{T}\right)g^{-1/6}_\text{eff}(T) \,,
\end{equation}
--and assuming a large enough reheating temperature $T_\text{b}\sim \sqrt{H_\text{b}\mplanck}$, 
$H_\text{b}\sim m_\phi$--  we find for a
moduli mass of $m_\phi \sim 100$ TeV a value of $\eta_\text{b}\sim 10^{-5} \seconds$ .
The resulting gravitational-wave background for $\eta_\text{e} \simeq \eta_\text{BBN}$ 
as expected from the usual moduli problem is 
indicated as the thick solid line in Fig.~\ref{fig:gwspectra}.
We have also indicated in Fig.~\ref{fig:gwspectra}
the results for moduli decaying much earlier, namely at $T^d_\phi \sim 10^6 \GeV$
(dashed line). 
Furthermore, for illustration we have indicated the 
suppression of the gravitational-wave spectrum in standard cosmology from a particle with $30 \TeV$ mass that enters
thermal equilibrium after inflation and decays before weakly-interacting massive particles (WIMP) freeze-out (thin solid line).
In standard cosmology such signals might not only arise from any extension of the standard model containing long-lived particles like the axino or modulinos, but could 
even be desired~\cite{Hasenkamp:2010if}.  
%

\begin{figure}[t]
\centering
\includegraphics[width=\columnwidth]{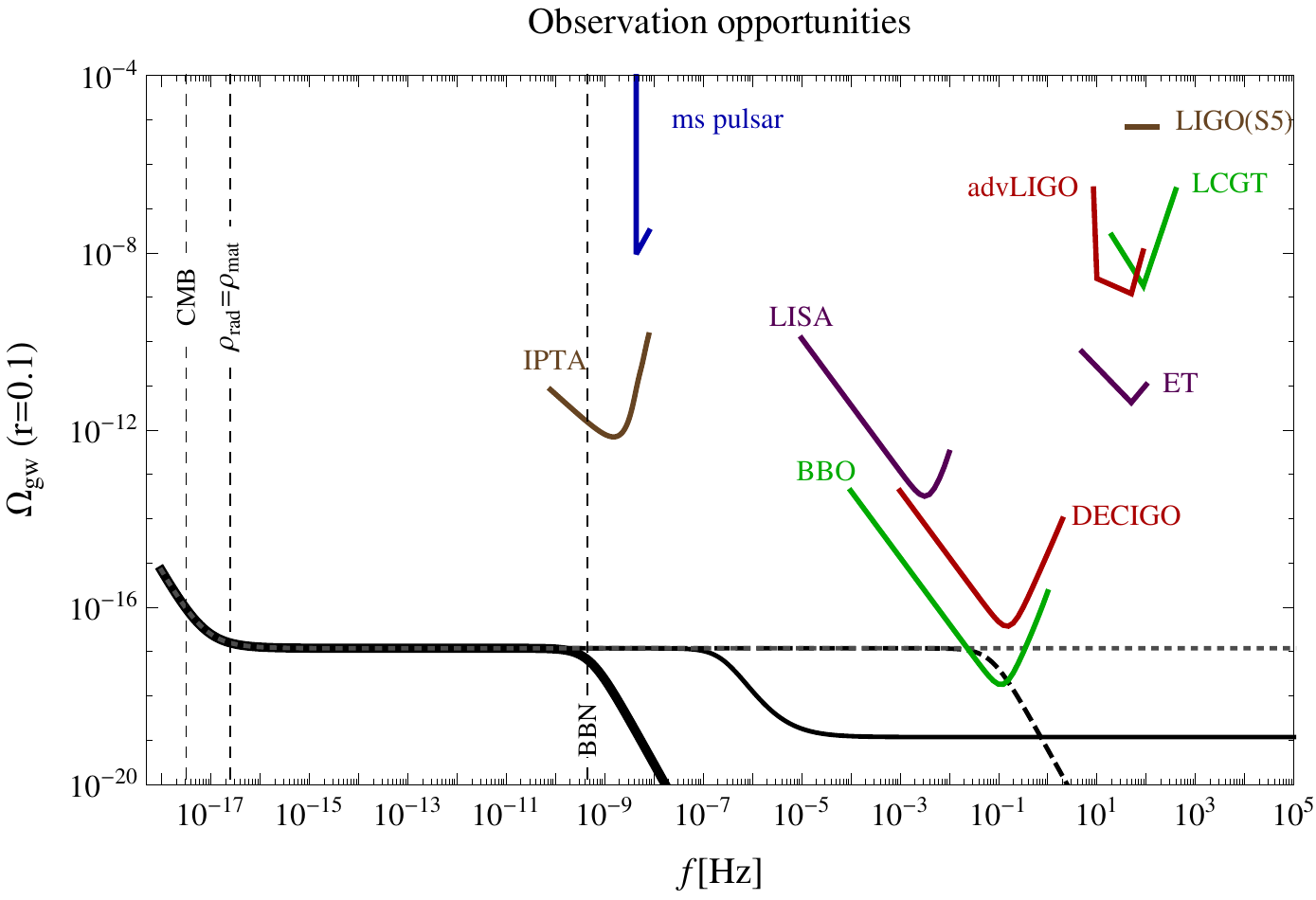} 
\caption{
Observation opportunities for the spectra of Fig.~\ref{fig:gwspectra}.
In addition to the spectra of Fig.~\ref{fig:gwspectra}, sensitivity curves of existing~\cite{existing}
 and future~\cite{future} gravitational-wave observatories are plotted.
}
\label{fig:gwdetection}
\end{figure}

We compare the spectra with present~\cite{existing} and future~\cite{future} gravitational-wave experiments in
Fig.~\ref{fig:gwdetection}; see footnotes and references therein for details on the experiments.
It is important that the decay temperature even of a much earlier modulus decay might be
inferred from observation. To compare to the signal of an intermediate reheating temperature see~\cite{Nakayama:2008wy}. 
For larger moduli masses the moduli dynamics might depend directly
on the reheating temperature. For the simplest potential, $V=m_\phi^2 \phi^2/2$,
we estimate $T_\text{R}\sim 10^9 \GeV$ as the minimal reheating temperature
corresponding to the sensitivity of BBO for our test to apply.

Interestingly,
pulsar timing arrays probe the scale of BBN, where the Universe is surely
radiation dominated. 
It would be particularly interesting, if they 
became sensitive to inflationary gravitational waves.
The present CMB limit on $r$ from~\cite{Dunkley:2010ge,Keisler:2011aw} is $r<0.2$ on CMB scales. 

\vspace{0.1cm}

\section{Conclusions and Outlook}
We have shown that the gravitational-wave background from inflation is strongly modified
by an intermediate matter dominated era. 
If string compactification leads to an intermediate matter dominated phase,
this will be observable in the gravitational-wave background: 
on CMB scales one will detect the unmodified background 
from inflation, e.g., with the Planck satellite~\cite{planck} or with a future 
CMB polarimeter~\cite{cmbpolarimeter}. However, on higher frequencies like milli-Hz or Hz probed
by the gravitational-wave detectors indicated in Fig.~\ref{fig:gwdetection} no signal will
be detected.
In other words, if these experiments will detect the signal from the inflationary
gravitational-wave background as expected from the CMB, this will rule out all
string compactifications that contain at least one scalar with a mass $ \lesssim 10^{12}\GeV$---corresponding
to the sensitivity of BBO---that
acquires a large initial oscillation amplitude after inflation 
and has only gravitational interaction strength.
A correspondingly high supersymmetry breaking scale, for example, of the order of the GUT scale 
may well render superstring models unobservable.

Even though our derivation presented here is for an intermediate matter dominated phase, 
the qualitative result remains true also for a phase of thermal 
inflation. Such a phase would dilute gravitational waves on sub-Hubble scales even more
strongly, and would render them undetectable for the experiments indicated in 
Fig.~\ref{fig:gwdetection}, while not affecting CMB scales.
A (possibly) observable gravitational-wave spectrum created after thermal inflation
were easily distinguishable from the primordial one by its shape~\cite{Easther:2008sx}.

Furthermore, from our derivation it is clear that for some other, non-inflationary, 
intermediate epoch starting at time $\eta_\text{b}$ and ending at $\eta_\text{e}$, with equation of
state $P=w\rho$ with $-1/3<w\le1$, the inflationary gravitational-wave spectrum will be
suppressed (or enhanced for $w>1/3$) by a factor $\alpha(k)$ with
\begin{eqnarray}
\Omega_\text{gw}^\text{final} &=& \alpha(k)\times \Omega_\text{gw}^\text{std} ~~ \mbox{ with } \nonumber \\  \alpha(k)&=& \left\{\begin{array}{ll}
1 & \mbox{if } k<\eta_\text{e}^{-1} \\
(k\eta_\text{e})^{2(3w-1)/(3w+1)} & \mbox{if } \eta_\text{e}^{-1}< k<\eta_\text{b}^{-1} \\
\left( \frac{\eta_\text{e}}{\eta_\text{b}}\right)^{2(3w-1)/(3w+1)} & \mbox{if } \eta_\text{e}^{-1}< k .
\end{array} \right.
\end{eqnarray}
Our fitting formula for $T(k)$ reproduces this behavior for an intermediate 
matter dominated era, $w=0$. For $w\neq 0$ the exponents  $\pm2$ in the 
denominator would have to be replaced by $\pm2(1-3w)/(1+3w)$.

Of course there is the caveat that (as also many other inflationary models) inflation
in models inspired by string theory~\cite{Baumann:2009ni}, depending on the compactification
and moduli stabilization, may predict only a very low
gravitational-wave background that cannot be measured by proposed experiments, neither
in the CMB nor directly on smaller scales. 
In this case, an experiment that would be able to
detect the background generated in the intermediate matter dominated 
era as predicted in~\cite{Durrer:2009zm} should be conceived.
Such a background is, however, suppressed with respect to the amplitude from inflation
by the ratio of the corresponding Hubble rates 
$(H_\text{b}/H_\text{inf})^2$, where $H_\text{b}$ 
denotes the Hubble rate at the beginning of the 
intermediate matter dominated  phase and $H_\text{inf}$ the one during inflation.

In conclusion, we have shown that combining future CMB polarization measurements
with very sensitive gravitational-wave probes such as  BBO can provide a crucial test for
a large class of string models. Since string theory is our best candidate for a fundamental 
theory of Nature,
and there has not been proposed any experiment to test it, it is 
of uttermost importance that we realize a BBO--like experiment.
\vspace{0.1cm}
\paragraph*{Acknowledgements:}
We thank Masaki Ando, Alessandra Buonanno, Daniel Figueroa, Kentaro Kuroda and 
Xavier Siemens for providing us information about the different experimental sensitivity 
curves and Alexander Westphal for helpful discussions. 
JH  is supported by the German Science Foundation (DFG) via the
Junior Research Group ``SUSY Phenomenology'' within the Collaborative
Research Center 676 ``Particles, Strings and the Early Universe''.
RD acknowledges support from the Swiss NSF and the French CNRS.

\end{document}